\documentclass[12pt]{iopartmod1}
\usepackage{color, graphicx}
\usepackage{bm}

\usepackage{iopams}

\font\smalltesti=cmmi8 at 8.0pt
\font\smallseveni=cmmi7 at 5.83pt
 1  

\begin{document}

\title[CR Structures and Twisting Vacuum Spacetimes: Type II and More Special]
{CR Structures and Twisting Vacuum Spacetimes with Two Killing
Vectors and Cosmological Constant: Type II and More Special}

\author{Xuefeng Zhang and Daniel Finley}

\address{Department of Physics and Astronomy, University of New Mexico, Albuquerque, NM 87131 USA}
\eads{\mailto{zxf@unm.edu} and \mailto{finley@phys.unm.edu}}

\begin{abstract}
Based on the CR formalism of algebraically special spacetimes by
Hill, Lewandowski and Nurowski, we derive a nonlinear system of two
real ODEs, of which the general solution determines a twisting type
II (or more special) vacuum spacetime with two Killing vectors
(commuting or not) and at most seven real parameters in addition to
the cosmological constant $\Lambda$. To demonstrate a broad range of
interesting spacetimes that these ODEs can capture, special
solutions of various Petrov types are presented and described as
they appear in this approach. They include Kerr-NUT, Kerr and
Debney/Demia\'{n}ski's type II, Lun's type II and III (subclasses of
Held-Robinson), MacCallum and Siklos' type III ($\Lambda<0$), and
the type N solutions ($\Lambda\neq 0$) we found in an earlier paper,
along with a new class of type II solutions as a nontrivial limit of
Kerr and Debney's type II solutions. Also, we discuss a situation in
which the two ODEs can be reduced to one. However, constructing the
general solution still remains an open problem.
\end{abstract}


\submitto{\CQG}

\maketitle


\section{Introduction}
\label{introduction}

All algebraically special Einstein spaces---vacuum but possibly with
a nonzero cosmological constant $\Lambda$---possess a repeated
principal null direction, which generates a foliation of the
spacetime by a 3-parameter congruence of shearfree and null
geodesics \cite{Goldberg62,Adamo12}. This 3-dimensional parameter
space can be identified with a 3-dimensional real manifold described
by the theory of CR structures with one complex and one real
coordinate. CR structures were first introduced by Poincar\'e and
extensively studied by E. Cartan \cite{Cartan32I,Cartan32II}. Good
sources of background on the relationship between spacetimes and CR
structures can be found, for instance, in the thesis of Nurowski
\cite{Nurowski93}. Recently, Hill, Lewandowski and Nurowski
\cite{Hill08} generalized earlier work of
\cite{Lewandowski90,Nurowski93} to provide a new formulation of
twisting algebraically special spacetimes with cosmological
constant. It allows a classification of algebraically special
spacetimes according to Cartan's classification of 3-dimensional CR
structures.

There is indeed a method to determine the equivalence of two
solutions of Einstein's equations without having to construct
explicit coordinate transformations that map one into the other, the
idea of which was originated in work of Cartan and pushed forward by
Brans \cite{Brans65}, Karlhede \cite{Karlhede80}, and Skea
\cite{Skea00a,Skea00b}. Cartan also created a method for determining
equivalence of two CR manifolds, which is much simpler than the
method mentioned above for 4-dimensional manifolds. Because of the
correspondence between the two, it allows a considerably simpler
approach to determine the equivalence of two twisting algebraically
special spacetimes, or, more usefully in this paper, the lack of
such an equivalence, thereby guaranteeing that two solutions are
distinct. In addition to this important reason for using CR
structures, it also provides a different formulation of Einstein's
equations, which exhibits certain invariant features that are
desirable for calculations, as compared to other formulations. It is
favorable to have this invariant approach to study the decomposition
of Einstein's equations into some manageable form, and to have one
that prefers non-zero values of the twist, since only a very limited
number of such spacetimes with non-zero twists, in Petrov types II,
III, and N, are actually available for study.

In the theory of exact solutions, Einstein's equations are usually
solved under assumptions of the existence of some symmetry group
\cite{Stephani03}, i.e., Killing vectors. For instance, Kerr and
Debney \cite{Kerr70} have determined all diverging (twisting or not)
algebraically special vacuums ($\Lambda=0$) with three or more
Killing vectors. The case with two Killing vectors, however, is
still not solved completely. We intend to address this problem in
this paper with the extension to include a nonzero cosmological
constant.

Building on the work of Hill, Lewandowski, and Nurowski
\cite{Hill08}, we first present, in Section 2, the twisting type II
vacuum metric formulated according to CR geometry, together with our
calculated Weyl scalars. Then in Section 3, we establish the
transformation from the CR formalism to the canonical frame that is
widely used, e.g., in \cite{Stephani03}. In Section 4, we generalize
the ansatz that was found for twisting type N solutions in
\cite{Zhang12a}, thereby reducing the field equations to only two
coupled real ODEs for two unknown functions of a single variable.
From Section 5 to 8, we show, by using the transformation in Section
3, that a large variety of previously-known, twisting solutions of
types II, III, as well as D (\cite{Stephani03}, Chapters 29 and 38),
with at least two Killing vectors, correspond to special solutions
of these ODEs. Moreover, we study a special case when the number of
ODEs can be reduced to one, which generalizes our previous results
on type N. Though the general solution is yet to be found, we
believe that this extension of ODEs is quite worthwhile and should
provide a start for future steps forward in the study of twisting
exact solutions.

\section{CR structures and the field equations}

A CR structure\footnote{All our considerations are local.} is a
3-dimensional real manifold $M$ equipped with an equivalence class
of pairs of 1-forms $\lambda$ (real) and $\mu$ (complex) satisfying
\begin{equation*}
 \lambda \wedge \mu \wedge \bar\mu \neq 0.
\end{equation*}
Another pair $(\lambda',\mu')$ is equivalent to $(\lambda,\mu)$, iff
there exist functions $f\neq 0$ (real) and $h\neq 0$, $g$ (complex)
on $M$ such that
\begin{equation*} 
 \lambda' = f\lambda, \qquad \mu' = h\mu + g\lambda, \qquad
 \bar\mu' = \bar h \bar\mu + \bar g \lambda.
\end{equation*}
For our purpose, we further assume \cite{Hill08,Cartan32I}
\begin{eqnarray}
 \mu = \rmd \zeta, \qquad \bar\mu  = \rmd \bar\zeta, \label{dmu} \\
 \rmd\lambda = \rmi\mu \wedge \bar\mu
 + (c\mu + \bar c \bar\mu )\wedge \lambda, \label{dlambda}
\end{eqnarray}
where $\zeta$ and $c$ are some complex-valued functions on $M$.
Taking the closure of \eref{dlambda}, we obtain a reality condition
on the derivatives of $c$:
\begin{equation}
 \partial\bar c =  \bar\partial c,
\end{equation}
The same function $c$ also appears in the commutation relations of
the dual basis of vector fields:
\begin{equation} \label{partials}
 \eqalign{
 \left(\partial_0, \partial, \bar\partial \right)
 \hbox{~~dual to~~} \left(\lambda, \mu, \bar\mu \right), \nonumber \\
 \left[\partial, \bar\partial\, \right] = -\rmi \partial_0, \qquad
 \left[\partial_0, \partial \right] = c \partial_0, \qquad
 \left[\partial_0, \bar\partial\, \right] = \bar c \partial_0.}
\end{equation}
There is then the following theorem telling us how to construct an
algebraically special spacetime on the basis of $M$.

\textbf{Theorem 1.} \cite{Hill08,Nurwoski08} The CR structure
(\ref{dmu}-\ref{partials}) on $M$ can be lifted to a spacetime
$\mathcal{M}=M\times \mathbb{R}$ equipped with the metric
\begin{equation} \label{metric}
 \eqalign{
 \mathbf{g} = 2 \left( \theta^1\theta^2 + \theta^3\theta^4 \right),
 \qquad \theta^1 = \mathcal{P}\,\mu = \bar\theta^2, \\
 \theta^3 = \mathcal{P}\,\lambda,
 \qquad \theta^4 = \mathcal{P} \left(\rmd r + \mathcal{W}\mu
 + \bar\mathcal{W} \bar\mu + \mathcal{H}\lambda \right), }
\end{equation}
where $\mathcal{P}\neq 0$, $\mathcal{H}$ (real) and $\mathcal{W}$
(complex) are arbitrary functions on $\mathcal{M}$. The spacetime
\eref{metric} admits a geodesic, shearfree and twisting null
congruence along the vector field $\partial_r$ ($r\in \mathbb{R}$),
of which the 3-parameter leaf spaces ($r=\mathrm{const.}$) have the
same CR structure as $M$. It further satisfies the Einstein equation
$\textrm{Ric}(\mathbf{g})=\Lambda \mathbf{g} $, iff the metric
components can be written as
\begin{eqnarray}
 \mathcal{P} = \frac{p}{\cos(\case{r}{2})}, \qquad \mathcal{W} = \rmi\,a\,(\rme^{-\rmi r} + 1), \\
 \mathcal{H} = \frac{n}{p^4}\, \rme^{2\rmi r} + \frac{\bar{n}}{p^4}\, \rme^{-2\rmi r}
 + q\,\rme^{\rmi r} + \bar q \,\rme^{-\rmi r} + h, \\
 a = c + 2 \partial \log p, \\
 q = \frac{3n+\bar{n}}{p^4} + \frac{2}{3} \Lambda p^2 + \frac{2 \partial p\, \bar\partial p
 - p\, \left(\partial \bar\partial p + \bar\partial \partial p \right)}{2 p^2} - \frac{\rmi}{2}\,
 \partial_0 \log p - \bar\partial c, \\
 h = 3\frac{n+\bar{n}}{p^4} + 2\Lambda p^2 + \frac{2 \partial p\, \bar\partial p
 - p\, \left(\partial \bar\partial p + \bar\partial \partial p \right)}{p^2} - 2 \bar\partial c,
\end{eqnarray}
where $c$, $n$ (complex) and $p$ (real), all functions on $M$
(independent of $r$), satisfy the following set of equations:
\begin{eqnarray}
 \partial \bar c = \bar\partial c, \label{ccb} \\
 \left[ \partial \bar\partial + \bar\partial \partial + \bar c
 \partial + c \bar\partial + \frac{1}{2} c \bar c + \frac{3}{4}
 \left(\partial \bar c + \bar\partial c \right) \right] p =
 \frac{n+\bar{n}}{p^3} + \frac{2}{3} \Lambda p^3, \label{NurowskiEqn} \\
 \partial n + 3c\,n = 0, \label{pncn} \\
 R_{33} = 0. \label{R330}
\end{eqnarray}
Here the Ricci tensor component $R_{33}$ as well as the Weyl scalars
$\Psi_2$, $\Psi_3$ and $\Psi_4$ are given by
\begin{eqnarray}
 R_{33} &= \Bigg\{ \frac{8}{p^4}\, (\partial + 2 c)
 \! \left[ p^2\! \left(\partial\bar{I} - 2\Lambda (2\bar\partial \log p + \bar c) p^2 \right) \right]
 \nonumber \\
 & + \frac{16}{p}\, \Lambda \Big[ \Big(\partial \bar\partial +
 \bar\partial \partial + \bar c \partial
 + c \bar\partial + \frac{1}{2} c \bar c + \frac{3}{4}
 (\partial \bar c + \bar\partial c ) \Big) p -
 \frac{n+\bar{n}}{p^3} - \frac{2}{3} \Lambda p^3 \Big]
 \nonumber \\
 & + \frac{16\rmi}{p^3}\, \partial_0\! \Big(\frac{n}{p^3}\Big)
 \Bigg\} \cos^4\! \left(\frac{r}{2} \right), \label{R33}
\end{eqnarray}
\begin{equation} \label{Psi2}
 \Psi_2 = \frac{n}{2p^6}\, (\rme^{\rmi r} + 1)^3,
\end{equation}
\begin{eqnarray}
 \Psi_3 =& \Bigg\{ \frac{2\rmi}{p^2} \left[\partial\bar{I}
 - 2\Lambda (2\bar\partial \log p + \bar c) p^2 \right] \label{Psi3} \\
 & + 6\rmi \left(2 \bar\partial \log p + \bar{c} \right)
 \frac{n}{p^6}\, (\rme^{2\rmi r} - 1)
 - 4\rmi \bar\partial \Big(\frac{n}{p^6}\Big) (\rme^{\rmi r} + 1)
 \Bigg\} \rme^{\rmi r/2} \cos^3\! \left(\frac{r}{2}\right),
 \nonumber
\end{eqnarray}
\begin{equation*}
 \fl \eqalign{
 \Psi_4 = 2 \rme^{-\rmi r/2} \cos^3 \left(\frac{r}{2}\right)
 \Bigg\{\! -(2\bar\partial \log p + \bar c) \left[\partial \bar{I}
 - 2\Lambda (2\bar\partial \log p + \bar c) p^2\right]
 \frac{\rme^{2\rmi r} - 1}{p^2}
 \\
 + (\bar\partial + 2\bar{c}) \left[\partial\bar{I} - 2\Lambda (2\bar\partial \log p + \bar c) p^2\right]
 \frac{\rme^{\rmi r} + 1}{p^2} + \frac{\rmi}{p^2} \partial_0 \bar{I}
 \\
 + \frac{2}{3} \Lambda \left[(\bar\partial + \bar c)
 (2\bar\partial \log p + \bar{c}) + 2(2\bar\partial \log p + \bar{c})^2 \right]
 -3(2\bar\partial \log p + \bar{c})^2 \frac{n}{p^6}\, \rme^{4\rmi r}
 \\
 + \left[ (\bar\partial - 2\bar\partial \log p)(2\bar\partial \log p + \bar{c})\, \frac{n}{p^6}
 + 3(2\bar\partial \log p + \bar{c}) \bar\partial \Big(\frac{n}{p^6}\Big) \right] \rme^{3\rmi r}
 \\
 + \left[ 3\left( 2\bar\partial^2 \log p - 16 (\bar\partial \log p)^2
 + 2\bar{c} \bar\partial \log p + \bar{c}^2 \right) \frac{n}{p^6}
 + \frac{(16\bar\partial \log p + \bar{c}) \bar\partial n - \bar\partial^2 n}{p^6} \right] \rme^{2\rmi r}
 \\
 + \left[ 3\left( 2 \bar\partial^2 \log p - 8(\bar\partial \log p)^2 + 8\bar{c} \bar\partial \log p
 - \bar\partial \bar{c} - \bar{c}^2\right) \frac{n}{p^6}
 + \frac{7(2\bar\partial \log p - \bar{c}) \bar\partial n - 2\bar\partial^2 n}{p^6} \right] \rme^{\rmi r}
 \\
 + 2\left( \bar\partial^2 \log p - 2(\bar\partial \log p)^2 + 5\bar{c} \bar\partial \log p
 - \bar\partial \bar{c} - 3\bar{c}^2\right) \frac{n}{p^6}
 + \frac{(4\bar\partial \log p - 5\bar{c}) \bar\partial n - \bar\partial^2 n}{p^6}
 \Bigg\}, }
\end{equation*}
with the function $I$ defined by
\begin{equation*}
 I = \partial \left(\partial \log p + c \right) + \left(\partial \log p + c \right)^2.
\end{equation*}

Following the same procedure as \cite{Hill08}, which uses Cartan's
structure equations to calculate the curvature tensor, we present
our calculated $R_{33}$, $\Psi_3$ and $\Psi_4$ above with nonzero
$\Lambda$ and $n$, as a complement to \cite{Hill08}. Moreover, to
facilitate future calculations, we have arranged the expression of
$R_{33}$ so that its second square bracket can be immediately
removed by the field equation \eref{NurowskiEqn}, whereas the terms
$\partial\bar{I} - 2\Lambda (2\bar\partial \log p + \bar c) p^2$ are
made prominent as they also appear in $\Psi_3$ and $\Psi_4$.

To solve the field equations (\ref{ccb}-\ref{R330}) in practice, one
needs to introduce a real coordinate system $(x, y ,u)$ on $M$ such
that
\begin{equation}
 \!\!\! \begin{array}{ll}
 \zeta = x + \rmi y, \qquad & \partial_\zeta = \frac{1}{2} \left(\partial_x - \rmi \partial_y \right), \\
 \partial = \partial_\zeta - L \partial_u,
 \qquad & \partial_0 = \rmi(\bar\partial L - \partial\bar{L}) \partial_u,
 \end{array} \qquad
 \lambda = \frac{\rmd u + L \rmd\zeta + \bar{L} \rmd\bar\zeta}{\rmi(\bar\partial L - \partial\bar{L})}, \label{lambda}
\end{equation}
with $L=L(\zeta, \bar\zeta, u)$ a complex-valued function
\cite{Hanges88} satisfying
\begin{equation}
 \bar\partial L - \partial\bar{L} \neq 0,
\end{equation}
which is needed for a nonzero twist (cf. \eref{compStph}). In
addition, $L$ relates to the function $c$ by
\begin{equation}
 c = -\partial \ln (\bar\partial L - \partial\bar{L}) - \partial_u L, \label{cL}
\end{equation}
as imposed by the commutation relations \eref{partials}. Hence
generally, the system (\ref{ccb}-\ref{R330}) are in fact PDEs for
the unknown functions $L$, $n$ and $p$ of the coordinate variables
$(\zeta,\bar\zeta,u)$.

For other possible coordinate choices, the metric
(\ref{dmu}-\ref{cL}) admits the following coordinate freedom
(\cite{Zhang12b}, see Section 2.6):
\begin{equation} \label{CoTfm}
 r'=r, \qquad \zeta' = f(\zeta),
 \qquad u'=F(\zeta,\bar\zeta,u), \qquad \partial_u F \neq 0,
\end{equation}
with $f(\zeta)$ holomorphic and $F$ a real-valued function, which
generates the transformation laws
\begin{eqnarray}
 \mu' = f'\mu, \qquad \lambda' = f'\bar{f}' \lambda,
 \qquad f' \equiv \rmd f/\rmd \zeta, \\
 \partial' = \frac{1}{f'} \partial,
 \qquad \partial_0' = \frac{1}{f'\bar{f}'} \partial_0,
 \qquad \left[\partial', \bar\partial' \, \right] = -\rmi \partial_0',
 \qquad \left[\partial_0', \partial' \right] = c' \partial_0', \\
 c' = \frac{1}{f'} c + \frac{f''}{(f')^2},
 \qquad p' = \frac{1}{|f'|} p,
 \qquad n' = \frac{1}{(f'\bar{f}')^3} n, \label{cpnpcpn} \\
 L' = -\frac{1}{f'} (\partial_\zeta F - L \partial_u F) = -\frac{1}{f'} \partial F,
 \quad \bar\partial'L' - \partial'\bar{L}'
 = \frac{\partial_u F}{f'\bar{f}'} (\bar\partial L - \partial \bar{L}).
\end{eqnarray}
Note that the function $F$ does not appear above on the level of
$c$, $p$ and $n$, as well as $\partial$ and $\partial_0$, the fact
of which indicates an invariant feature of the CR formalism. As
expected, the field equations for the new $p'$, $c'$ and $n'$ take
on the same form of (\ref{ccb}-\ref{R330}) with
$(\partial_0,\partial,\bar\partial)$ simply replaced by
$(\partial_0',\partial',\bar\partial')$. These transformation
properties will be used to simplify our metrics.

\section{Transformations to the canonical frame}
\label{sec:canonical}

Given the algebraically special twisting metric form
(\ref{dmu}-\ref{cL}) formulated according to CR geometry, it is
important to know how it is related to other pre-existing formalisms
that have been extensively studied in the past. Here we quote from
\cite{Stephani03} (p. 439--441) a most commonly used one by Kerr,
Debney \emph{et al.} \cite{Kerr63,Debney69,Robinson69}. For
simplicity, we only consider $\Lambda=0$ and follow closely the
notation of \cite{Stephani03} with sub- or superscript $s$ added to
avoid confusion.

\textbf{Theorem 2.} A spacetime admits a geodesic, shearfree and
twisting null congruence along the vector field $\partial_{r_s}$ and
satisfies the Einstein equation $\textrm{Ric}(\mathbf{g})=0$, iff
the metric can be written as
\begin{equation} \label{metricStph}
 \eqalign{
 \mathbf{g} = 2 (\mathbf{\omega}^1 \omega^2 - \omega^3 \omega^4),
 \qquad \omega^1 = -\frac{\rmd\zeta}{P_s \bar\rho_s} = \bar\omega^2, \\
 \omega^3 = \rmd u + L\rmd \zeta + \bar{L} \rmd \bar\zeta,
 \qquad \omega^4 = \rmd r_s + W_s \rmd \zeta + \bar{W}_s \rmd \bar\zeta + H_s \omega^3,}
\end{equation}
with metric components
\begin{equation} \label{compStph}
 \eqalign{
 \rho_s^{-1} = -(r_s + \rmi \Sigma_s),
 \qquad \frac{2\rmi \Sigma_s}{P_s^2} = \bar\partial L - \partial \bar{L} \neq 0, \\
 W_s = \rho_s^{-1} \partial_u L + \rmi \partial\Sigma_s, \qquad \partial = \partial_\zeta - L\partial_u, \\
 H_s = \frac{1}{2}K_s - r_s\partial_u \log P_s - \frac{m_s r_s + M_s \Sigma}{r_s^2 + \Sigma^2}, \\
 K_s = 2P_s^2 \mathrm{Re} \left[\partial \left(\bar\partial \log P_s - \partial_u \bar{L} \right)\right]}
\end{equation}
such that the functions $m_s$, $M_s$, $P_s$ (real) and $L$
(complex), all only dependent on the coordinates $(\zeta, \bar\zeta,
u)$, satisfy a system of PDEs:
\begin{eqnarray}
 P_s^{-3} M_s = \mathrm{Im}\ \partial\partial \bar\partial \bar\partial V_s,
 \qquad P_s = \partial_u V_s, \label{VsPs}\\
 \partial (m_s + \rmi M_s) = 3 (m_s + \rmi M_s) \partial_u L, \\
 \partial_u \left[ P_s^{-3}(m_s + \rmi M_s)\right]
 = P_s [\partial + 2(\partial \log P_s - \partial_u L)] \partial I_s,
 \label{Rs330}
\end{eqnarray}
where the function $I_s$ is defined by
\begin{equation} \label{Is}
 I_s = \bar\partial \left(\bar\partial \log P_s - \partial_u \bar{L} \right)
 + \left(\bar\partial \log P_s - \partial_u \bar{L} \right)^2 = P_s^{-1} \partial_u \bar\partial \bar\partial V_s.
\end{equation}
Additionally, the Weyl scalar $\Psi^s_2$ is given by
\begin{equation*}
 \Psi^s_2 = (m_s + \rmi M_s) \rho_s^3.
\end{equation*}

In this metric form, the coordinates $(\zeta, \bar\zeta, u)$ and the
function $L$ have been chosen identically with those introduced in
\eref{lambda}; hence, each is not given a sub- or superscript $s$.
Taking $\Lambda=0$ and by a tedious but straightforward calculation,
one can show that the metrics (\ref{metricStph}-\ref{Is}) and
(\ref{dmu}-\ref{cL}) are equivalent to each other by the
transformation \cite{Zhang12b}
\begin{eqnarray}
 & P_s = \frac{2 p}{\rmi(\bar\partial L - \partial \bar{L})}, \label{Psp} \\
 & r_s = -\frac{2 p^2}{\rmi(\bar\partial L - \partial \bar{L})} \tan\left(\frac{r}{2}\right),
 \qquad |r|<\pi, \label{rsr} \\
 & m_s = \frac{16(n - \bar{n})}{(\bar\partial L - \partial \bar{L})^3},
 \qquad M_s = \frac{16(n + \bar{n})}{\rmi(\bar\partial L - \partial \bar{L})^3}, \label{mMsn}
\end{eqnarray}
with the inverse
\begin{eqnarray}
 p = \frac{\rmi}{2} (\bar\partial L - \partial \bar{L}) P_s, \label{pPs} \\
 r = 2\arctan\left( -\frac{2}{\rmi (\bar\partial L - \partial \bar{L}) P_s^2}\, r_s \right),
 \label{rrs} \\
 n = \frac{1}{32}(m_s + \rmi M_s)(\bar\partial L - \partial \bar{L})^3. \label{nmMs}
\end{eqnarray}
In particular, the field equation \eref{VsPs} can be transformed
into \eref{NurowskiEqn} with $\Lambda=0$, despite their drastically
different appearances. Also, one gets $I_s=\bar{I}$ when
substituting \eref{Psp} into the definition \eref{Is}. For more
details about these transformations, one may see Section 2.6 of
\cite{Zhang12b}. The relation \eref{pPs} and \eref{nmMs} together
with \eref{cL} will be used later to translate known solutions of
the canonical field equations (\ref{VsPs}-\ref{Rs330}) to solutions
of (\ref{ccb}-\ref{R330}).

\section{Reductions to ODEs}

Now we go back to the metric (\ref{dmu}-\ref{cL}). Following the
same idea as \cite{Zhang12a} for solving the field equations, we
assume that the unknowns $p$, $c$ and $n$ have no $u$-dependence,
i.e., $\partial_0 p = \partial_0 c =
\partial_0 n = 0$. This assumption avoids the involvement of the
function $L$ inside the operator $\partial$ since now we have, e.g.,
$\partial p = \partial_\zeta p$. Therefore the system
(\ref{ccb}-\ref{R330}) becomes effectively PDEs for the unknowns
$c$, $p$ and $n$ instead of $L$, $p$ and $n$. Once
$c=c(\zeta,\bar\zeta)$ is solved, one may further determine a
function $L$ without $u$-dependence from \eref{cL} (cf. \eref{Lxz}).
Altogether, this means that the resulting spacetime shall possess a
Killing vector in the $u$-direction, which is, in fact, an
assumption widely used in many research articles on algebraically
special solutions (see, e.g., \cite{Stephani03} Chapter 29).

We apply the assumption and rewrite the system
(\ref{ccb}-\ref{R330}) (likewise for $\Psi_3$ and $\Psi_4$) as
\begin{eqnarray}
 \partial_\zeta \bar{c} = \partial_{\bar\zeta} c, \label{ccb1} \\
 2 \partial_\zeta \partial_{\bar\zeta}p + \bar{c} \partial_\zeta p
 + c \partial_{\bar\zeta} p + \frac{1}{2} c \bar{c} p + \frac{3}{2} (\partial_\zeta \bar{c}) p
 = \frac{n+\bar{n}}{p^3} + \frac{2}{3}\Lambda p^3, \label{NurowskiEqn1} \\
  \partial_\zeta n + 3c\,n = 0, \label{pncn1} \\
 (\partial_\zeta + 2 c)\! \left[ p^2 \partial_\zeta \bar{I}
 - 2\Lambda (2\partial_{\bar\zeta} \log p + \bar{c}) p^4 \right] = 0, \label{R3301}
\end{eqnarray}
where in the last equation ($R_{33}=0$) we have used
\eref{NurowskiEqn} to simplify the expression of $R_{33}$, and the
function $I$ is given by
\begin{equation*}
 I = \partial_\zeta \left(\partial_\zeta \log p + c \right) + \left(\partial_\zeta \log p + c \right)^2.
\end{equation*}
This is the set of PDEs we aim to solve for the unknowns
$p(\zeta,\bar\zeta)$, $c(\zeta,\bar\zeta)$ and $n(\zeta,\bar\zeta)$.

Generalizing the ansatz \cite{Zhang12a,Zhang12b} we found from the
classical symmetries \cite{Krasilshchik99} of the type N case of
(\ref{ccb1}-\ref{R3301}) with $n=0$, we assume the following forms
for the unknowns:
\begin{equation}  \label{pcnz}
 \fl p(\zeta, \bar\zeta) = \frac{F_1(z)}{\sqrt{A \bar{A}}}, \quad
 c(\zeta, \bar\zeta) = \frac{\partial_\zeta A + \rmi F_2(z) + C_1}{A}, \quad
 n(\zeta, \bar\zeta) = \frac{F_3(z) + \rmi F_4(z)}{(A\bar{A})^3}
\end{equation}
with a new real variable
\begin{equation*}
 z = -\rmi \left( \int \frac{1}{A} \rmd \zeta -
 \int \frac{1}{\bar A} \rmd \bar{\zeta} \right)
 = \mathrm{Im} \int \frac{2}{A} \rmd \zeta.
\end{equation*}
Here the function $A=A(\zeta)$ is an arbitrary function of $\zeta$
that is sufficiently smooth, and the constant $C_1$ and the
undetermined functions $F_{1-4}(z)$ are all real-valued. The
constraint equation \eref{ccb1} has been taken into account in the
form of $c(\zeta,\bar\zeta)$ so that it is satisfied.

Inserting the ansatz \eref{pcnz} into
(\ref{NurowskiEqn1}-\ref{R3301}), we obtain a remarkable reduction
to a system of four compatible real ODEs for $F_{1-4}(z)$ only, with
all other dependence on $A, \bar{A}\neq 0$ factored out:
\begin{eqnarray}
 0 = -F_1'' + F_2 F_1' + \frac{1}{3}\Lambda F_1^3
 + \frac{1}{4} (3 F_2' - F_2^2 - C_1^2) F_1 + \frac{F_3}{F_1^3}, \label{NurowskiEqnz} \\
 0 = F_3' - 3(F_2 F_3 + C_1 F_4), \label{F3ODE} \\
 0 = F_4' + 3(C_1 F_3 - F_2 F_4), \label{F4ODE} \\
 0 = (H' - 2F_2 H)' - 2 F_2(H' - 2F_2 H) + 4 C_1^2 H, \label{R330z}
\end{eqnarray}
where the function $H(z)$ is defined by
\begin{equation*}
 H = F_1'' F_1 - (F_1')^2 - \Lambda F_1^4 - F_2' F_1^2.
\end{equation*}
The two inner equations are derived from the real and imaginary
parts of \eref{pncn1}, respectively. Note that they are linear in
$F_3$ and $F_4$. Thus if $F_2$ is given, one can solve them by
\begin{equation} \label{F34F2}
 \eqalign{
 F_3 = \exp\left(3\! \int\! F_2\rmd z \right)
 [B_1 \sin(3C_1 z) - B_2 \cos(3C_1 z)], \\
 F_4 = \exp\left(3\! \int\! F_2\rmd z \right)
 [B_2 \sin(3C_1 z) + B_1 \cos(3C_1 z)], }
\end{equation}
where $B_{1,2}$ are real constants. With $F_3$ expressed in terms of
$F_2$, we are left with only two nonlinear equations for $F_1$ and
$F_2$:
\begin{equation} \label{2ODEs}
 \eqalign{
 0 =& -F_1'' + F_2 F_1' + \frac{1}{3}\Lambda F_1^3 + \frac{1}{4} (3 F_2' - F_2^2 - C_1^2) F_1 \\
 &+ \frac{\exp\left(3\! \int\! F_2\rmd z \right)}{F_1^3}\, [B_1 \sin(3C_1 z) - B_2 \cos(3C_1 z)], \\
 0 =& (H' - 2F_2 H)' - 2 F_2(H' - 2F_2 H) + 4 C_1^2 H, }
\end{equation}
which can be easily converted to a set of ODEs if one introduces,
e.g., $F_2=K'(z)$ to remove the integral in the first equation. The
system \eref{2ODEs} with \eref{F34F2}, or alternatively
(\ref{NurowskiEqnz}-\ref{R330z}), constitutes the main result of
this paper.

Before moving on to solve (\ref{NurowskiEqnz}-\ref{R330z}) in the
next few sections, we should make a few general remarks concerning
the metric (\ref{dmu}-\ref{cL}) equipped with the ansatz
\eref{pcnz}. First, despite the appearance of a free function
$A(\zeta)\neq 0$ in the ansatz, its different choices do not
generate new metrics. In fact, using the coordinate change
$\zeta'=\int\! \frac{2}{A(\zeta)} \rmd\zeta$ permitted by
\eref{CoTfm} and, accordingly, the transformation law \eref{cpnpcpn}
with $f'=2/A(\zeta)$, one can always replace a function $A$ by a
constant $A=2$. This is also consistent with the fact that the local
CR structure determined by the function $c$ in \eref{pcnz} is
independent of the choice of $A(\zeta)$ \cite{Zhang12a}. Hence for
simplicity, we can just set $A(\zeta)=\bar{A}(\bar\zeta)=2$ without
loss of generality, and hence $z=\mathrm{Im}\zeta=y$. With this
choice of $A$, the function $L$ and the 1-form $\lambda$ can be
determined from \eref{cL} and \eref{lambda} as
\begin{eqnarray}
 L = -\rme^{-C_1 x} \int \exp\left(\int F_2 \rmd z \right) \rmd z, \label{Lxz} \\
 \lambda = \frac{\rme^{C_1 x} \rmd u
 - 2 \left[\int \exp\left(\int F_2 \rmd z \right) \rmd z \right] \rmd x}
 {\exp\left(\int F_2 \rmd z \right)}. \label{lambdaz}
\end{eqnarray}
Once again, though there exist other $L$'s satisfying \eref{cL}, one
can always use the remaining coordinate freedom
$u'=F(\zeta,\bar\zeta,u)$ (cf. \eref{CoTfm}) to convert them to the
$u$-independent real expression \eref{Lxz} \cite{Zhang12a}. Given
such $L$ and $\lambda$, the class of metrics determined by
(\ref{pcnz}-\ref{R330z}) admits at least two Killing vectors
\begin{equation} \label{KV}
 X_1 = \partial_u, \qquad X_2 = \partial_x - C_1 u\, \partial_u,
\end{equation}
with the commutation relation
\begin{equation*}
 [X_1, X_2] = -C_1 X_1.
\end{equation*}
These vectors, verifiable by direct calculation, are both inherited
from the symmetries of the underlying CR structures
\cite{Nurowski88,Zhang12b}.

\section{Type D solutions: Kerr-NUT}

All type D vacuum solutions, twisting or not, are known
\cite{Kinnersley69,Plebanski76}, which include perhaps the most
famous algebraically special solutions such as Kerr's rotating
black-hole solution. Hence it is worthwhile to consider whether our
equations (\ref{NurowskiEqnz}-\ref{R330z}) can capture some of these
physically important solutions. But first we should comment that the
CR formalism (\ref{dmu}-\ref{cL}) used here is constructed on just
one single shearfree null congruence aligned with a multiple
principal null direction , which is unique in type II, III and N
spacetimes. However, a type D spacetime possesses two such
congruences, each along one of the two doubly degenerate principal
null directions, and consequently one cannot treat them at the same
time in the CR formalism. This suggests that the CR formalism may
not provide the most convenient approach for finding type D
solutions, as compared to other approaches that are specially
designed to make use of both congruences.

A spacetime is of type D iff it satisfies the conditions
\begin{equation} \label{CTD}
 3\Psi_3\Psi_4 - 2\Psi_2^2 = 0, \qquad \Psi_2\neq 0.
\end{equation}
This equality with the ansatz \eref{pcnz} applied gives rise to a
number of lengthy ODEs (as one may sense by looking at the
expression of $\Psi_4$) from the coefficients of various powers of
$\rme^{\rmi r}$ required to vanish. This fairly complicated
situation (except when $\Psi_3=\Psi_4=0$, see \ref{App:TD}) needs a
specialized paper to elaborate; hence it is not further discussed
here (also because there is no new type D solution to be found).
Instead, by applying the results of Section \ref{sec:canonical}, we
will simply show that the Kerr-NUT solution can be retrieved as a
special solution of (\ref{NurowskiEqnz}-\ref{R330z}) through the
ansatz \eref{pcnz}.

In the canonical frame (\ref{metricStph}-\ref{Is}), the Kerr-NUT
solution \cite{Stephani03} (p. 453) is given by
\begin{equation*}
 \eqalign{
 P_s = 1 + \frac{\zeta \bar\zeta}{2},
 \qquad L = -\frac{\rmi}{\zeta P_s^2} [ 2M + (M + a)\zeta \bar\zeta ], \\
 m_s = m, \qquad M_s = M, \qquad \Lambda = 0,}
\end{equation*}
where $m$, $M$ and $a$, each a real constant, are called the mass,
the NUT parameter and the Kerr parameter, respectively. Then
inserting them into \eref{pPs}, \eref{cL} and \eref{nmMs}, we obtain
the following solution to (\ref{ccb1}-\ref{R3301}):
\begin{eqnarray*}
 p = -\frac{4(M-a)+2(M+a)\zeta\bar\zeta}{(2+\zeta\bar\zeta)^2}, \\
 c = \frac{2\bar\zeta [2(M-2a)+(M+a)\zeta\bar\zeta]}
 {[2(M-a)+(M+a)\zeta\bar\zeta](2+\zeta\bar\zeta)}, \\
 n = -\frac{16\rmi [2(M-a)+(M+a)\zeta\bar\zeta]^3 (m+\rmi M)}{(2+\zeta\bar\zeta)^9}.
\end{eqnarray*}
which, as expected, satisfies the condition \eref{CTD} for type D
with nonzero $\Psi_3$ and $\Psi_4$. Without a dependence on $u$,
these expressions can be cast into the form of our ansatz
\eref{pcnz} by
\begin{eqnarray*}
 A(\zeta) = -\rmi \zeta, \qquad z = \log(\zeta \bar\zeta), \qquad C_1=0,
\end{eqnarray*}
such that
\begin{equation} \label{KerrNUT}
 \eqalign{
 \fl F_1 = -\frac{2[2(M-a)+(M+a)\rme^z]\rme^{z/2}}{(2+\rme^z)^2},
 \qquad F_2 = \frac{4(M-a)+8a\rme^z-(M+a)\rme^{2z}}
 {[2(M-a)+(M+a)\rme^z](2+\rme^z)}, \\
 \fl F_3 = \frac{16M[2(M-a)+(M+a)\rme^z]^3\rme^{3z}}{(2+\rme^z)^9},
 \qquad F_4 =
 -\frac{16m[2(M-a)+(M+a)\rme^z]^3\rme^{3z}}{(2+\rme^z)^9}.}
\end{equation}
One can verify that they are indeed a solution to
(\ref{NurowskiEqnz}-\ref{R330z}) with $\Lambda=0$. Note that the
function $A(\zeta)$ above will still be serving as a free function,
as long as $F_{1-4}$ are obtained. The Kerr-NUT solution is
contained in the Demia\'{n}ski solution as a special case (cf.
\eref{Demianski}). For other examples of type D solutions, see
\ref{App:TDCS} and \eref{new} with $B_2=0$.

\section{Type N solutions with nonzero cosmological constant}

The type N solutions require
\begin{equation*}
 \Psi_2=\Psi_3=0, \qquad \Psi_4\neq 0.
\end{equation*}
These conditions lead to the following special case of
(\ref{NurowskiEqnz}-\ref{R330z}) with $F_3 = F_4 = 0$:
\begin{equation} \label{TN}
 \eqalign{
 0 = -F_1^{\prime \prime} + F_2 F_1' + \case{1}{3}\Lambda F_1^3
 + \case{1}{4} (3 F_2' - F_2^2 - C_1^2) F_1, \nonumber \\
 \Psi_3 \propto H' - 2F_2 H - 2\rmi C_1 H = 0,}
\end{equation}
plus one inequality
\begin{eqnarray*}
 \Psi_4 &=& -\frac{4\Lambda}{3\bar{A}^2 F_1^2}\,
 \Big[ 2 F_1 F_1'' + 6 (F_1')^2 - 10 (F_2 + \rmi C_1) F_1 F_1' \nonumber \\
 & & - (F_2' - 3 F_2^2 - 6 \rmi C_1 F_2 + 3 C_1^2) F_1^2 \Big]\,
 \rme^{-\rmi r/2} \cos^3\! \left(\frac{r}{2}\right) \neq 0,
\end{eqnarray*}
which imposes $\Lambda\neq 0$ for type N. To better see that the
system \eref{TN} is included in (\ref{NurowskiEqnz}-\ref{R330z}),
one can rewrite equation \eref{R330z} as
\begin{equation*}
 0 = (H' - 2F_2 H - 2\rmi C_1 H)'
 - 2(F_2' - \rmi C_1)(H' - 2F_2 H - 2\rmi C_1 H).
\end{equation*}
In fact, the field equation $R_{33}=0$ can always be removed by
$\Psi_2=\Psi_3=0$ for general type N vacuums (cf.
(\ref{R33}-\ref{Psi3})). Notice that the second equation of
\eref{TN} is complex; we have two cases for solutions.

\emph{Case 1:} $H=0$, $\Lambda\neq 0$. This simpler case has been
investigated in \cite{Zhang12a} (see \cite{Zhang12b} for more
details). The equations for this case read
\begin{equation} \label{TN1}
 \eqalign{
 0 = -F_1^{\prime \prime} + F_2 F_1' + \case{1}{3}\Lambda F_1^3
 + \case{1}{4} (3 F_2' - F_2^2 - C_1^2) F_1, \\
 0 = H = -F_1^{\prime \prime} F_1 + (F_1')^2 + \Lambda F_1^4 + F_2' F_1^2.}
\end{equation}
By introducing a real function $J=J(z)$ and
\begin{equation} \label{F12J}
 F_1 = \pm \sqrt{J'}, \qquad F_2 = \frac{J^{\prime\prime}}{2 J'} - \Lambda J, \qquad J'>0,
\end{equation}
we can reduce the first equation of \eref{TN1} to
\begin{equation} \label{JEQ}
 J''' = \frac{(J^{\prime\prime})^2}{2J'} - 2 \Lambda J J^{\prime\prime} -
 \frac{10}{3} \Lambda (J')^2 - 2 (\Lambda^2 J^2 + C_1^2) J',
\end{equation}
while the second equation is automatically satisfied. Since this ODE
has no explicit dependence on the variable $z$, we can immediately
lower its order by the transformation $J'=P(J)>0$ such that
\begin{equation} \label{PEQ}
 P'' = - \frac{(P' + 2 \Lambda J)^2}{2P} - \frac{2 C_1^2}{P} - \frac{10}{3} \Lambda,
 \qquad \Lambda\neq 0,
\end{equation}
which, in the case of $C_1=0$, can be further reduced to an Abel ODE
\cite{Polyanin95} of the first kind
\begin{equation} \label{Abel}
 f' = \frac{4}{t} \left(t+\frac{3}{2}\right)\left(t+\frac{1}{3}\right) f^3
 + \frac{5}{t} \left(t+\frac{2}{5}\right) f^2 + \frac{1}{2t}\ f,
\end{equation}
by $J = \exp(\int\! f(t) \rmd t)/\Lambda$, $P(J) = t \exp(2 \int\!
f(t) \rmd t)/\Lambda$. Unfortunately, this Abel ODE has not been
identified as a known solvable type.

Various aspects of the equation \eref{PEQ} were examined in
\cite{Zhang12a}, including the weak Painlev\'{e} property
\cite{Conte08} and constructions of various special and series
solutions. All degenerate solutions of type O ($\Psi_4=0$,
conformally flat) were found. The only known type N solution with $
\Lambda\neq 0$ in closed forms was first discovered by Leroy
\cite{Leroy70} and presented in the CR formalism by Nurowski
\cite{Nurwoski08}. It corresponds to
\begin{equation} \label{Leroy}
 \eqalign{
 P(J) = -\frac{1}{3}\Lambda J^2 - \frac{3 C_1^2}{4\Lambda} >0, \\
 F_1 = \pm \frac{\sqrt{3} C_1}{2 s \sin(\frac{1}{2} C_1 (z + C_0))},
 \qquad F_2 = -\frac{2 C_1}{\tan(\frac{1}{2} C_1 (z + C_0))},}
\end{equation}
with $\Lambda=-s^2<0$ and $C_0$ a real constant (removable by a
translation $z+C_0\rightarrow z$). Nonetheless, the equation
\eref{PEQ} does also admit type N solutions with $\Lambda>0$ and two
additional parameters besides $\Lambda$ and $C_1$.

\emph{Case 2:} $H\neq 0$, $C_1=0$, $\Lambda\neq 0$. The system
\eref{TN} becomes
\begin{equation} \label{TN2}
 \eqalign{
 0 = -F_1^{\prime \prime} + F_2 F_1' + \case{1}{3}\Lambda F_1^3
 + \case{1}{4} (3 F_2' - F_2^2) F_1, \\
 0 = H' - 2F_2 H.}
\end{equation}
Very little is known about the solutions of this system except for
one of type O \cite{Zhang12b} given by
\begin{equation} \label{hyqN}
 F_1 = \pm \frac{\sqrt{6}}{2s(z+C_0)}, \qquad F_2 = -\frac{2}{z+C_0},
 \qquad H = \frac{3}{4 s^2 (z+C_0)^4},
\end{equation}
with $\Lambda = -s^2<0$ and $C_0$ a real constant. Particularly,
this solution has the hyperquadric CR structure (the most
symmetrical one) \cite{Jacobowitz90}.

\section{Type III solutions}

Similar to the case of type N, the equations for type III
($\Psi_2=0$, $\Psi_3\neq 0$; $F_3 = F_4 = 0$) are given by
\begin{equation} \label{TIII}
 \eqalign{
 0 = -F_1'' + F_2 F_1' + \case{1}{3}\Lambda F_1^3
 + \case{1}{4} (3 F_2' - F_2^2 - C_1^2) F_1, \\
 0 = (H' - 2F_2 H)' - 2 F_2(H' - 2F_2 H) + 4 C_1^2 H,}
\end{equation}
which are subject to
\begin{equation*}
 \Psi_3 \propto H' - 2F_2 H - 2\rmi C_1 H \neq 0.
\end{equation*}
Using the first equation of \eref{TIII}, we can lower the order of
the second ODE, such that the resulting set of equations contains
derivatives up to the second-order in $F_1$ and the third-order in
$F_2$. Hence the general solution carries another five real
parameters in addition to $\Lambda$ and $C_1$. Considering that one
of these parameters is simply the translation $z \rightarrow z+C_0$
(no explicit dependence on $z$ in (\ref{NurowskiEqnz}-\ref{R330z})),
and thus removable, one can see that the final type III metric
determined by \eref{TIII} has at most six parameters including
$\Lambda$ and $C_1$ (see the conclusions).

Two classes of twisting type III vacuum solutions are known,
respectively for $\Lambda=0$ and $\Lambda<0$. The one with
$\Lambda=0$ is due to Held \cite{Held74} and Robinson
\cite{Robinson75}, which generalizes the non-twisting
Robinsion-Trautman type III vacuum solution and generally admits
only one Killing vector $\partial_u$. The subclasses with two
Killing vectors (commuting or not) were found by Lun \cite{Lun78}.
It can be shown that Lun's case I type III metric corresponds to the
following solution of \eref{TIII}:
\begin{equation} \label{Lun3}
 \eqalign{
 \Lambda =0, \qquad C_1 = 0, \\
 F_1 = \frac{\sqrt{3}}{2 z^3} (E_1 z^{2+\sqrt{13}/2} + E_2 z^{2-\sqrt{13}/2}), \\
 F_2 = -\frac{(5-\sqrt{13}) E_1 z^{2+\sqrt{13}/2} + (5+\sqrt{13}) E_2 z^{2-\sqrt{13}/2}}
 {2z (E_1 z^{2+\sqrt{13}/2} + E_2 z^{2-\sqrt{13}/2})}. }
\end{equation}
Likewise, his case II type III metric puts forth a second solution
of \eref{TIII}:
\begin{equation} \label{Lun31}
 \eqalign{
 \Lambda =0, \qquad C_1 = \frac{1}{4}, \\
 F_1 = \frac{\sqrt{3}}{4\cos(\frac{z}{2})} (E_1 G + E_2 G^{-1}), \\
 F_2 = \frac{\sqrt{13}\, (E_1 G - E_2 G^{-1})}{4\cos(\frac{z}{2})\, (E_1 G + E_2 G^{-1})}
 + \frac{5}{4}\tan(\case{z}{2}), \\
 G(z) = \left( \frac{\sin(\case{z}{2})+1}
 {\cos(\case{z}{2})} \right)^{\sqrt{13}/2}. }
\end{equation}
In both cases, $E_{1,2}$ are real constants. More details will be
given in the next section as degenerate cases of the related type II
solutions \eref{Lun2} and \eref{Lun21}.

The other known type III solution requires $\Lambda<0$ and is due to
MacCallum and Siklos \cite{Siklos81} (see also \cite{Stephani03} p.
201). As a solution of \eref{TIII}, it is given by
\begin{equation} \label{Siklos}
 F_1 = \pm \frac{\sqrt{39}}{4s(z+C_0)}, \qquad F_2 = -\frac{5}{2(z+C_0)},
 \qquad C_1=0,
\end{equation}
with $\Lambda=-s^2$, a real constant $C_0$ and
\begin{equation*}
 \Psi_3 \propto H'-2F_2 H = -\frac{585}{256\Lambda(z+C_0)^5}.
\end{equation*}

\section{Type II solutions}

Based on the structure of (\ref{NurowskiEqnz}-\ref{R330z}), we can
consider the type II solution ($\Psi_2\neq 0$, $3\Psi_2 \Psi_4 -
2\Psi_3^2\neq 0$) according to three different cases.

\emph{Case 1:} $F_3=0$, $F_4\neq 0 \Rightarrow C_1=0$. The equations
(\ref{NurowskiEqnz}-\ref{R330z}) are reduced to
\begin{eqnarray*}
 0 = -F_1'' + F_2 F_1' + \case{1}{3}\Lambda F_1^3
 + \case{1}{4} (3 F_2' - F_2^2) F_1, \\
 0 = (H' - 2F_2 H)' - 2 F_2(H' - 2F_2 H), \\
 F_4 = B_1 \exp\left(3\! \int\! F_2\rmd z \right),
\end{eqnarray*}
with $B_1\neq 0$ a real constant. Since the first two equations
above are identical to \eref{TIII} with $C_1=0$ (also cf. \eref{TN1}
and \eref{TN2}), one can generate this kind of type II solutions
directly from existing type N and III solutions, i.e., \eref{Leroy}
with $C_1=0$, \eref{Lun3} and \eref{Siklos}, which works as if one
is adding a ``mass source'' to them. A similar idea can be found in
\cite{Stephani03} p. 447. In addition, for type D solutions with
$F_3=0$ and $C_1=0$, see \ref{App:TDCS}.

\emph{Case 2:} $F_3\neq 0$, $C_1=0$. The associated equations are
given by
\begin{eqnarray*}
 0 = -F_1'' + F_2 F_1' + \case{1}{3}\Lambda F_1^3
 + \case{1}{4} (3 F_2' - F_2^2) F_1 + \frac{F_3}{F_1^3}, \\
 0 = (H' - 2F_2 H)' - 2 F_2(H' - 2F_2 H), \\
 F_2 = \frac{F_3'}{3F_3}, \qquad
 F_4 = B_1 \exp\left(3\! \int\! F_2\rmd z \right) = B_1 F_3,
\end{eqnarray*}
with $B_1$ a real constant. Certainly one may use the third equation
above to turn the first two into ODEs for $F_1$ and $F_3$ only.

Lun's case I solution \cite{Lun78,McIntosh87} with four parameters
can be shown to belong to this case. It reads, in the canonical
frame (\ref{metricStph}-\ref{Is}),
\begin{eqnarray*}
 P_s = \sqrt{\case{2}{3}}\, (\zeta + \bar\zeta)^{3/2}, \qquad \zeta = x + \rmi y, \\
 L = -\frac{3\rmi}{16} x^{-3/2} \left[(3+\sqrt{13}) E_1 x^{\sqrt{13}/2}
 + (3-\sqrt{13}) E_2 x^{-\sqrt{13}/2} \right] + \frac{3\rmi M}{32 x^3}, \\
 m_s = m, \qquad M_s = M, \qquad \Lambda = 0,
\end{eqnarray*}
or, in the CR formalism as a solution of (\ref{ccb1}-\ref{R3301}) or
(\ref{NurowskiEqnz}-\ref{R330z}),
\begin{equation} \label{Lun2}
 \eqalign{
 A(\zeta) = -2\rmi, \qquad z = x, \qquad C_1 = 0, \\
 p = \frac{1}{2} F_1(z) = \frac{\sqrt{3}}{16 z^3}\, G(z), \\
 c = -\frac{1}{2} F_2(z) \\
 \ \ =(z G)^{-1} \left[(5-\sqrt{13}) E_1 z^{2+\sqrt{13}/2}
 + (5+\sqrt{13}) E_2 z^{2-\sqrt{13}/2} + 6M z^{1/2} \right], \\
 n = \frac{1}{64} \left(F_3(z) + \rmi F_4(z)\right)
 = \frac{27\, \rmi}{2^{20} z^{27/2}}\, G^3 (m+\rmi M), \\
 G(z) = 4\left( E_1 z^{2+\sqrt{13}/2} + E_2 z^{2-\sqrt{13}/2} \right) + 3M z^{1/2}, }
\end{equation}
with $m$, $M$ and $E_{1,2}$ real constants. When $m = M = 0$, the
solution degenerates to the type III solution \eref{Lun3}.

The Kerr and Debney/Demia\'{n}ski's four-parameter solution
\cite{Stephani03} (see p. 449) also falls under this case. It is
given by
\begin{equation*}
 \eqalign{
 P_s = 1 + \frac{\zeta \bar\zeta}{2},
 \qquad L = -\rmi P_s^2 \left[ 2M/\zeta + (M+a)\bar\zeta
 + \case{1}{4} b \bar\zeta \log(\bar\zeta/\sqrt{2}\,) \right], \\
 m_s = m, \qquad M_s = M, \qquad \Lambda = 0,}
\end{equation*}
and corresponds to
\begin{equation} \label{Demianski}
 \eqalign{
 A(\zeta) = -\rmi\zeta, \qquad z = \log(\zeta \bar\zeta),
 \qquad C_1=0, \\
 F_1 = -\frac{[16(M-a)+8(M+a)\rme^z + b G_1]
 \rme^{z/2}}{4(2+\rme^z)^2}, \\
 F_2 = \frac{8[4(M-a)+8a\rme^z-(M+a)\rme^{2z}] + b G_2}
 {8[2(M-a)+(M+a)\rme^z](2+\rme^z) + b G_1 (2+\rme^z)}, \\
 F_3 = \frac{M[16(M-a)+8(M+a)\rme^z + b G_1]^3\rme^{3z}}{32(2+\rme^z)^9}, \\
 F_4 = -\frac{m[16(M-a)+8(M+a)\rme^z + b G_1]^3\rme^{3z}}{32(2+\rme^z)^9}, \\
 G_1(z) = (2+\ln 2-z)(2-\rme^z) - 8, \\
 G_2(z) = (3+\ln 2-z)(4+\rme^{2z}) - 8(\ln 2-z)\rme^z - 24,}
\end{equation}
which is a solution of (\ref{NurowskiEqnz}-\ref{R330z}). Here $m$,
$M$, $a$ and $b$ are four real parameters. Clearly, the Kerr-NUT
solution \eref{KerrNUT} is a special case with $b=0$.

Besides these known solutions, we have obtained an additional one
(see the derivation in Case 3 below) which turns out to be a
limiting case ($C_1\rightarrow 0$) of the solution \eref{newKD}:
\begin{equation} \label{new}
 \eqalign{
 C_1 = 0, \qquad \Lambda = 0, \\
 F_1 = -2B_2 z^2 + C_2 z + C_3,
 \qquad F_2 = \frac{4B_2 z - C_2}{2B_2 z^2 - C_2 z - C_3}, \\
 F_3 = -B_2 F_1^3, \qquad F_4 = B_1 F_1^3.}
\end{equation}
Here $C_{2,3}$ and $B_{1,2}$ are real constants. Its comparisons
with Lun's and Demia\'{n}ski's solutions will be discussed in
\ref{App:Cmp}. Particularly when $B_2=0$, the solution becomes type
D.

\emph{Case 3:} $F_3\neq 0$, $C_1\neq 0 \Rightarrow F_4\neq 0$. This
corresponds to the most general case for solutions. As one may
check, Lun's case II four-parameter solution
\cite{Lun78,McIntosh87}, which is given by
\begin{eqnarray*}
 P_s = \sqrt{\case{2}{3}}\, (\zeta + \bar\zeta)^{3/2},
 \qquad \zeta = x + \rmi y, \qquad w=y/x, \\
 L = x^{-3/2} \Big\{ \case{1}{6} [
 E_1 (w+(w^2+1)^{1/2})^{\sqrt{13}/2} (w-\case{\sqrt{13}}{2}(w^2+1)^{1/2}) \\
 + E_2 (w+(w^2+1)^{1/2})^{-\sqrt{13}/2} (w+\case{\sqrt{13}}{2}(w^2+1)^{1/2}) \Big] \\
 + \case{3}{160} [(m+\rmi M)(1+\rmi w)^{3/2} (2-3\rmi w)
 + (m-\rmi M)(1-\rmi w)^{3/2} (2+3\rmi w)] \Big\}, \\
 m_s - \rmi M_s = (m+\rmi M) x^{3/2} (1+\rmi w)^{3/2}, \qquad \Lambda = 0,
\end{eqnarray*}
can be converted to a solution of (\ref{NurowskiEqnz}-\ref{R330z}):
\begin{equation} \label{Lun21}
 \fl \eqalign{
 A(\zeta)=\zeta, \qquad \zeta=|\zeta|\,\rme^{\rmi z/2}, \qquad C_1 = \frac{1}{4}, \\
 F_1 = \frac{\sqrt{3} (E_1 G + E_2 G^{-1})}{4\cos(\case{z}{2})}
 + \frac{3\sqrt{3}\,\sin(\case{z}{2})
 \left(M \sin(\case{z}{4}) - m \cos(\case{z}{4}) \right) }
 {16\cos^{5/2}(\case{z}{2})}, \\
 F_2 = \frac{4\cos^{3/2}(\case{z}{2}) [ 5\sin(\case{z}{2}) (E_1 G + E_2 G^{-1})
 + \sqrt{13} (E_1 G - E_2 G^{-1}) ] - 3 (M G_1 - m G_2) }
 {16 \cos^{5/2}(\case{z}{2}) (E_1 G + E_2 G^{-1})
 + 6 \sin(z) \left(M \sin(\case{z}{4}) - m \cos(\case{z}{4}) \right)}, \\
 F_3 = \frac{3\sqrt{3} F_1^3 \left( M \cos(\case{3z}{4}) + m \sin(\case{3z}{4}) \right)}
 {2^8 \cos^{9/2}(\case{z}{2})}, \qquad
 F_4 = \frac{3\sqrt{3} F_1^3 \left( -M \sin(\case{3z}{4}) + m \cos(\case{3z}{4}) \right)}
 {2^8 \cos^{9/2}(\case{z}{2})}, \\
 G(z) = \left( \frac{\sin(\case{z}{2})+1}{\cos(\case{z}{2})} \right)^{\sqrt{13}/2},
 \qquad G_1(z) = \case{5}{4} \sin(\case{5z}{4})
 - \case{7}{4} \sin(\case{3z}{4}) - 5\sin(\case{z}{4}), \\
 G_2(z) = \case{5}{4} \cos(\case{5z}{4})
 + \case{7}{4} \cos(\case{3z}{4}) - 5\cos(\case{z}{4}), }
\end{equation}
with $E_{1,2}$, $M$ and $m$ real constants. In the case of $M=m=0$,
the solution reduces to the type III solution \eref{Lun31}.

Besides Lun's example, we have also considered the special case of
$H=0$ for the system (\ref{NurowskiEqnz}-\ref{R330z}), which turns
out to be fully soluble when $\Lambda=0$. The derivation follows
closely the type N \emph{Case 1} (cf. \eref{TN1}), and utilizes the
same ansatz \eref{F12J} that makes the function $H(z)$ vanish
(hence, \eref{R330z} satisfied). More specifically, we have
\begin{equation} \label{newF}
 \eqalign{
 F_1 = \pm \sqrt{J'}, \qquad F_2 = \frac{J''}{2J'} - \Lambda J
 = \frac{F_1'}{F_1} - \Lambda J, \qquad J'>0, \\
 F_3 = F_1^3 \exp\left(-3\Lambda\! \int\! J\rmd z \right) [B_1 \sin(3C_1 z) - B_2 \cos(3C_1 z)], \\
 F_4 = F_1^3 \exp\left(-3\Lambda\! \int\! J\rmd z \right) [B_2 \sin(3C_1 z) + B_1 \cos(3C_1 z)], }
\end{equation}
with the last two equations derived from \eref{F34F2}. Therefore
when $\Lambda$ vanishes, the equation \eref{NurowskiEqnz} can be
reduced to a linear ODE for $F_1(z)$ alone:
\begin{equation*}
 F_1'' = -C_1^2 F_1 + 4[B_1 \sin(3C_1 z) - B_2 \cos(3C_1 z)],
 \qquad \Lambda = 0,
\end{equation*}
which has the general solution
\begin{equation} \label{newKD}
 \fl F_1 = C_2 \cos(C_1 (z+C_0))
 - \frac{1}{2C_1^2} [B_1 \sin(3C_1 z) - B_2 \cos(3C_1 z)],
 \qquad C_1\neq 0,
\end{equation}
or, if $C_1$ vanishes (cf. \eref{new}),
\begin{equation} \label{new1}
 F_1 = -2B_2 z^2 + C_3 z + C_4, \qquad C_1=0,
\end{equation}
where $C_{0-4}$ and $B_{1,2}$ are real constants. The conditions
$H=0$ and $\Lambda=0$ exclude type III and N as special cases. The
solution corresponding to \eref{newKD} with $C_1=-\case{1}{2}$
coincides with a special case of Kerr and Debney's type II solution
\cite{Stephani03} (p. 608), and hence is not new (see \ref{App:KD}
for more details). Also, by setting $C_0=0$,
$C_2=-\frac{B_2}{2C_1^2}+C_4$ and $B_1=-\frac{2}{3}C_1 C_3$, one can
obtain \eref{new1} from $\eref{newKD}$ in the limit $C_1\rightarrow
0$, whereas the two Killing vectors \eref{KV} become commuting. For
a generally non-vanishing $\Lambda$, we obtain from
\eref{NurowskiEqnz} an equation for $J(z)$:
\begin{equation} \label{JEQII}
 \eqalign{
 J''' &= \frac{(J^{\prime\prime})^2}{2J'} - 2 \Lambda J J^{\prime\prime} -
 \frac{10}{3} \Lambda (J')^2 - 2 (\Lambda^2 J^2 + C_1^2) J' \\
 & + 8 (J')^{1/2} \exp\left(-3\Lambda\! \int\! J\rmd z \right)
 [B_1 \sin(3C_1 z) - B_2 \cos(3C_1 z)],}
\end{equation}
which generalizes the equation \eref{JEQ} for type II. By
introducing $J=K'$ (or $-3\Lambda J=K'/K$, etc.), one can transform
this equation to a fourth-order ODE for $K(z)$. However, we do not
have any type II solution with $\Lambda\neq 0$ for this ODE.

\section{Conclusions}

With the real coordinates $\{x, z, u, r\}$ and $A(\zeta)=2$ in the
ansatz \eref{pcnz}, we present here our new class of vacuum twisting
type II metrics admitting two Killing vectors:
\begin{equation*}
 \mathbf{g} = \frac{F_1^2(z)}{2 \cos^2(\frac{r}{2})}
 \left[ \rmd\zeta \rmd\bar\zeta + \lambda \left(\rmd r
 + \mathcal{W} \rmd\zeta + \bar\mathcal{W} \rmd\bar\zeta + \mathcal{H} \lambda \right) \right],
\end{equation*}
with $\zeta = x + \rmi z$ and
\begin{eqnarray*}
 \lambda = \frac{\rme^{C_1 x} \rmd u
 - 2 \left[\int \exp\left(\int F_2 \rmd z \right) \rmd z \right] \rmd x}
 {\exp\left(\int F_2 \rmd z \right)}, \\
 \mathcal{W} = \frac{1}{2} \left( \frac{2F_1'}{F_1} - F_2 + \rmi C_1 \right) (\rme^{-\rmi r}+1), \\
 \fl \mathcal{H} = -\frac{1}{2} \left[ \left(\frac{F_1'}{F_1}\right)^\prime
 - \Lambda F_1^2 - F_2' - \frac{2( F_3(\cos r + 1) + F_4\sin r )}{F_1^4}
 \right]\! (\cos r + 1) - \frac{1}{6} \Lambda F_1^2 \cos r,
\end{eqnarray*}
where $C_1$ is an arbitrary real parameter and the real functions
$F_{1-4}(z)$ satisfy
\begin{equation} \label{FODEs}
 \eqalign{
 0 = -F_1'' + F_2 F_1' + \frac{1}{3}\Lambda F_1^3
 + \frac{1}{4} (3 F_2' - F_2^2 - C_1^2) F_1 + \frac{F_3}{F_1^3}, \\
 0 = (H' - 2F_2 H)' - 2 F_2(H' - 2F_2 H) + 4 C_1^2 H, \\
 F_3 = \exp\left(3\! \int\! F_2\rmd z \right)
 [B_1 \sin(3C_1 z) - B_2 \cos(3C_1 z)], \\
 F_4 = \exp\left(3\! \int\! F_2\rmd z \right)
 [B_2 \sin(3C_1 z) + B_1 \cos(3C_1 z)],}
\end{equation}
with the function $H(z)$ defined by
\begin{equation*}
 H = F_1^{\prime \prime} F_1 - (F_1')^2 - \Lambda F_1^4 - F_2' F_1^2.
\end{equation*}
One can determine $F_{1,2}$ from the first two equations of
\eref{FODEs}, which are in fact a pair of ODEs for $F_1$ and $K=\int
F_2\rmd z$. This allows the metric to have at most eight real
parameters including $\Lambda$ and $C_1$ (see discussion in Section
7), and its two Killing vectors are
\begin{equation*}
 \partial_u, \qquad \partial_x - C_1 u\, \partial_u.
\end{equation*}
Most of the previously known twisting vacuum solutions with two
Killing vectors, as presented in Chapters 29 and 38 of
\cite{Stephani03}, have been shown to belong to this class, the only
exception being the general case of Kerr and Debney's type II
solution (see \ref{App:KD}). Additionally, for $H=0$, the system
\eref{FODEs} can be reduced to a single ODE \eref{JEQII}, or even be
fully integrated when $\Lambda=0$. This leads to the discovery of a
limiting solution \eref{new} (type II with two commuting Killing
vectors) of a special case of Kerr and Debney's type II solution,
which we believe has not been discussed or published before. Despite
all these special solutions with maximally four parameters, the
general solution of \eref{FODEs} is still quite unknown. We believe
that this problem poses a major challenge.

Altogether, we hope that this work may provide a platform for all
types of twisting algebraically special solutions to be studied in a
connected and unified manner, given the history that many of those
known solutions were derived by quite different approaches or
special assumptions. For future research, this new class of metrics
may be further examined to study important issues such as the cosmic
no-hair conjecture, the asymptotic stability of the Kerr solution
\cite{Natorf12}, and the formation of rotating black holes that
might be described by certain solutions of \eref{FODEs} (see, e.g.,
\cite{Bicak95,Oliveira04} and references therein).



\appendix

\section{Type D solutions: $\Psi_3=\Psi_4=0$}
\label{App:TD}

For simplicity, we consider type D solutions with $\Psi_2\neq 0$ and
$\Psi_3=\Psi_4=0$. In order to acquire $\Psi_3=0$ with $\Psi_3$
depending on $r$, we need, at least, for the coefficient of
$\rme^{2\rmi r}$ to vanish in \eref{Psi3},
\begin{equation} \label{TD}
 0 = 2\bar\partial \log p + \bar{c},
\end{equation}
which, through the ansatz \eref{pcnz}, adds two more equations to
the system:
\begin{equation} \label{TD1}
 C_1=0, \qquad F_1' = \case{1}{2} F_2 F_1.
\end{equation}
Thus we can simplify the original (\ref{NurowskiEqnz}-\ref{R330z})
to
\begin{eqnarray}
 F_2' = -\frac{4}{3}\Lambda F_1^2 - \frac{4F_3}{F_1^4}, \\
 F_3' = 3F_2 F_3, \\
 F_4' = 3F_2 F_4, \label{TD4}
\end{eqnarray}
with the last equation \eref{R330z} being automatically satisfied by
the equations presented above. The system (\ref{TD1}-\ref{TD4}) can
be fully solved with the general solution
\begin{equation} \label{TDsol}
 \eqalign{
 F_1 = C_3 \sec\! \Big(\frac{z + C_0}{C_2}\Big),
 \qquad F_2 = \frac{2}{C_2} \tan\! \Big(\frac{z + C_0}{C_2}\Big), \\
 \fl F_3 = -\frac{C_3^4(3 + 2\Lambda C_2^2 C_3^2)}{6C_2^2}
 \sec^6\! \Big(\frac{z + C_0}{C_2}\Big),
 \qquad F_4 = C_4 \sec^6\! \Big(\frac{z + C_0}{C_2}\Big).}
\end{equation}
Remarkably, though we have only started with one extra condition
\eref{TD}, it is enough for the solution \eref{TDsol} to be of type
D, i.e., that we have
\begin{equation*}
 \Psi_2 = \frac{-C_3^4(3+2\Lambda C_2^2 C_3^2)+ 6\rmi C_2^2 C_4}
 {12 C_2^2 C_3^6} (\rme^{\rmi r} + 1)^3,
 \qquad \Psi_3 = \Psi_4 = 0.
\end{equation*}

With $A(\zeta)=2$ and \eref{TDsol}, the resulting metric can be
written as
\begin{equation*}
 \mathbf{g} = \frac{C_4^2}{2 \cos^2(\frac{r}{2}) \cos^2(\frac{z + C_0}{C_2})}
 \left[ \rmd\zeta \rmd\bar\zeta + \lambda \left(\rmd r + \mathcal{H} \lambda \right) \right]
\end{equation*}
with $\zeta = x + \rmi z$ and
\begin{eqnarray*}
 \lambda = \cos^2\! \Big(\frac{z + C_0}{C_2}\Big) \left[ \rmd u
 - 2 C_2 \tan\! \Big(\frac{z + C_0}{C_2}\Big) \rmd x \right], \\
 \mathcal{H} = -\frac{1}{\cos^2(\frac{z + C_0}{C_2})}
 \left[ \cos^2\! \left(\frac{r}{2}\right) \!
 \left(\frac{\cos r}{C_2^2} + \frac{2 C_4 \sin r}{C_3^4} \right)
 + \frac{1}{6} C_3^2\Lambda (\cos 2r + 2\cos r) \right].
\end{eqnarray*}
One can immediately remove the parameter $C_0$ by $z + C_0
\rightarrow z$. Besides the cosmological constant $\Lambda$, the
metric contains three real parameters $C_2$, $C_3$ and $C_4$.

\section{Type D solutions from classical symmetries}
\label{App:TDCS}

Here we list a number of special solutions one may encounter when
searching for group-invariant solutions from classical symmetries
\cite{Krasilshchik99} of the system (\ref{NurowskiEqnz}-\ref{R330z})
or its various special cases. Incidentally, all these solutions turn
out to be of type D, even though the condition
$2\Psi_2^2=3\Psi_3\Psi_4$ is never used in their derivation. As
expected for type D solutions, they all have $C_1=0$, which means
that the two Killing vectors \eref{KV} are commuting. In what
follows, $C_{2-4}$ are real constants.

We start with solutions with $F_3=0$. For $\Lambda = 0$, we have
\begin{eqnarray*}
 F_1 = C_3 \rme^{C_2 z}, \qquad F_2 = 2 C_2,
 \qquad F_3 = 0, \qquad F_4 = C_4 \rme^{6C_2 z}, \\
 F_1 = \frac{C_2}{z^2}, \qquad F_2 = -\frac{3}{z},
 \qquad F_3 = 0, \qquad F_4 = \frac{C_3}{z^9}.
\end{eqnarray*}
For $\Lambda = -s^2<0$, we have (\cite{Zhang12a} and cf.
\eref{hyqN})
\begin{eqnarray*}
 F_1 = \frac{\sqrt{6}}{3s z}, \qquad F_2 = -\frac{5}{3z},
 \qquad F_3 = 0, \qquad F_4 = \frac{C_2}{z^5}, \\
 F_1 = \frac{\sqrt{6}}{2s z}, \qquad F_2 = -\frac{2}{z},
 \qquad F_3 = 0, \qquad F_4 = \frac{C_2}{z^6}.
\end{eqnarray*}
In case of $F_4=0$, they all become type O solutions of \eref{TN} or
\eref{TIII}.

For solutions that admit non-vanishing $F_3$ and $\Lambda$, we have
\begin{eqnarray*}
 F_1 = C_2\neq 0, \qquad F_2 = 0,
 \qquad F_3 = -\case{1}{3} C_2^6\Lambda, \qquad F_4 = C_3, \\
 F_1 = \frac{C_2}{z}, \qquad F_2 = -\frac{2}{z},
 \qquad F_3 = -\frac{3C_2^4 + 2C_2^6 \Lambda}{6z^6},
 \qquad F_4 = \frac{C_3}{z^6},
\end{eqnarray*}
Note that the first solution above is given by constants, which is a
consequence of the translational invariance ($z\rightarrow z+C_0$)
of the system (\ref{NurowskiEqnz}-\ref{R330z}). Lastly for
$\Lambda=0$, we obtain
\begin{equation*}
 F_1 = C_3 \rme^{C_2 z}, \qquad F_2 = \case{4}{3}C_2,
 \qquad F_3 = \case{1}{9} C_3^4 C_2^2 \rme^{4C_2 z},
 \qquad F_4 = C_4 \rme^{4C_2 z}.
\end{equation*}


\section{Comparisons of type II solutions with $C_1=0$}
\label{App:Cmp}

Here we compare the three type II solutions from \emph{Case 2},
i.e., \eref{Lun2}, \eref{Demianski} and \eref{new}, and show that
\eref{new} is different from the other two. Generally, to see that
two twisting type II vacuum metrics are different, i.e., not being
related by a coordinate transformation, it is sufficient to show
that their CR structures along the shearfree null congruences are
not equivalent \cite{Hill08}. This can be decided by evaluating the
six Cartan invariants \cite{Nurowski93,Zhang12a}, which are the same
only for two equivalent CR structures. Among these invariants, the
first one, in Cartan's original notation, is given by
\begin{equation*}
 \eqalign{
 \alpha(\zeta,\bar\zeta) = -\frac{5 \bar{r} \partial_\zeta r + r \partial_\zeta \bar{r} + 8 c r\bar{r}}
 {8 \sqrt{\bar{r}} \cdot \sqrt[8]{(r\bar{r})^{7}}}, \\
 \bar{r} = \case{1}{6}\left(\partial_{\zeta} l + 2 c l\right), \qquad
 l = -\partial_\zeta \partial_{\bar\zeta} c - c \partial_{\bar\zeta} c,}
\end{equation*}
which only relies on the function $c=c(\zeta,\bar\zeta)$ (hence
$F_2(z)$ from the ansatz \eref{pcnz}). The use of $\alpha$ alone
will be adequate for our comparison.

For simplicity, we consider the special case of \eref{new} with
$C_2=C_3=0$, i.e.,
\begin{equation*}
 F_1 = -2B_2 z^2, \qquad F_2 = \frac{2}{z},
 \qquad F_3 = -B_2 F_1^3, \qquad F_4 = B_1 F_1^3.
\end{equation*}
This solution is still of type II but has a constant invariant
$\alpha$ given by
\begin{equation*}
 \alpha^2 = -\frac{25}{14} \sqrt{21}.
\end{equation*}
Yet another case with $\alpha$ being constant is when $B_2=0$ in
\eref{new}, in which case the solution is of type D and
\begin{equation*}
 \alpha^2 = -\frac{16}{5} \sqrt{10}.
\end{equation*}
This same quantity $\alpha$ calculated from Lun's solution
\eref{Lun2}, however, is generally a function of $z$ and only
becomes a constant when two of the three free parameters $E_{1,2}$
and $M$ vanish, i.e., that we have
\begin{eqnarray*}
 \alpha^2 = \frac{\sqrt{15}}{10} \ \ \mathrm{for}\ \ E_1=E_2=0, \\
 \alpha^2 = \sqrt{2} (4-\sqrt{13}) \ \ \mathrm{for}\ \ E_1=M=0, \\
 \alpha^2 = \sqrt{2} (4+\sqrt{13}) \ \ \mathrm{for}\ \ E_2=M=0,
\end{eqnarray*}
none of which is equal to those of \eref{new}. As for
Demia\'{n}ski's solution \eref{Demianski}, one can see (using Maple)
that its invariant $\alpha$ is never a constant (even when $b=0$ for
type D; particularly for the NUT solution with $b=a=0$, $\alpha$ is
not defined due to $\bar{r}=0$, and the corresponding CR structure is
hyperquadric \cite{Lewandowski90}) within the full range of the
parameters $M$, $a$ and $b$. Therefore we conclude that the solution
\eref{new} is different from \eref{Lun2} and \eref{Demianski}.

\section{Kerr and Debney's type II solution}
\label{App:KD}

The solution by Kerr and Debney \cite{Kerr70} (see also
\cite{Stephani03} p. 608) admits two non-commuting Killing vectors
and reads
\begin{equation} \label{KDII}
 \eqalign{
 P_s = 1, \qquad L = A_1 \bar\zeta^2 \zeta^{1+\sigma}
 + A_2 \bar\zeta \zeta^{\sigma/3}, \\
 \rmi M_s - m_s = 2A_1(1+\sigma)\zeta^\sigma, \qquad \Lambda=0,}
\end{equation}
with $\mathrm{Re}\,\sigma = -3$ and $A_{1,2}$ complex constants. The
special case with $\sigma = -3$ can be captured by the ansatz
\eref{pcnz}, and it corresponds to ($E_{1-4}$ real)
\begin{equation*}
 \eqalign{
 A(\zeta)=\zeta, \qquad \zeta=|\zeta|\,\rme^{\rmi z/2}, \qquad C_1 = -\frac{1}{2}, \\
 F_1 = E_3\sin(\case{z}{2}) - E_4\cos(\case{z}{2})
 + 2E_1\sin(\case{3z}{2}) - 2E_2\cos(\case{3z}{2}), \\
 F_2 = F_1'/F_1, \qquad
 F_3 = -F_1^3 \left( E_1 \sin(\case{3z}{2}) - E_2 \cos(\case{3z}{2}) \right), \\
 F_4 = F_1^3 \left( E_2 \sin(\case{3z}{2}) + E_1 \cos(\case{3z}{2}) \right), \\
 A_1 = E_1 + \rmi E_2, \qquad A_2 = E_3 + \rmi E_4. }
\end{equation*}
Modulo some redefinition of parameters, this is the same solution as
\eref{newKD} and \eref{newF} with $C_1=-\case{1}{2}$. It is not
clear how to make such a conversion for the general case of
\eref{KDII}.


\end{document}